\documentclass{article}

\usepackage{microtype}
\usepackage{graphicx}
\usepackage{subfigure}
\usepackage{booktabs} % for professional tables

\usepackage{hyperref}

\usepackage[accepted]{icml2023}

\usepackage{amsmath}
\usepackage{amssymb}
\usepackage{mathtools}
\usepackage{amsthm}
\usepackage{textcomp}
\usepackage{stfloats}
\usepackage{url}
\usepackage{verbatim}
\usepackage{graphicx}
\usepackage{tikz}
\usepackage{comment}
\usepackage{color}
\usepackage{epsfig}
\usepackage{booktabs} % for professional tables
\usepackage{soul}
\usepackage{xcolor, relsize}
\usepackage{capt-of}
\usepackage{multirow}
\usepackage{tablefootnote}
\usepackage{pifont, tabularx}
\DeclareMathOperator*{\argmin}{argmin}

\usepackage[capitalize,noabbrev]{cleveref}

\theoremstyle{plain}

\theoremstyle{definition}

\theoremstyle{remark}

\usepackage[textsize=tiny]{todonotes}

\icmltitlerunning{Image Shortcut Squeezing}

\begin{document}

\twocolumn[
\icmltitle{Image Shortcut Squeezing:\\Countering Perturbative Availability Poisons with Compression}

% It is OKAY to include author information, even for blind
% submissions: the style file will automatically remove it for you
% unless you've provided the [accepted] option to the icml2023
% package.

% List of affiliations: The first argument should be a (short)
% identifier you will use later to specify author affiliations
% Academic affiliations should list Department, University, City, Region, Country
% Industry affiliations should list Company, City, Region, Country

% You can specify symbols, otherwise they are numbered in order.
% Ideally, you should not use this facility. Affiliations will be numbered
% in order of appearance and this is the preferred way.

\begin{icmlauthorlist}
\icmlauthor{Zhuoran Liu}{yyy}
\icmlauthor{Zhengyu Zhao}{xjtu,comp}
\icmlauthor{Martha Larson}{yyy}
%\icmlauthor{}{sch}
%\icmlauthor{}{sch}
\end{icmlauthorlist}

\icmlaffiliation{yyy}{Radboud University, Nijmegen, Netherlands}
\icmlaffiliation{xjtu}{Xi'an Jiaotong University, Xi'an, China}
\icmlaffiliation{comp}{CISPA Helmholtz Center for Information Security, Saarbr{\"u}cken, Germany}

\icmlcorrespondingauthor{Zhuoran Liu}{z.liu@cs.ru.nl}
\icmlcorrespondingauthor{Zhengyu Zhao}{zhengyu.zhao@cispa.de}
\icmlcorrespondingauthor{Martha Larson}{m.larson@cs.ru.nl}

% You may provide any keywords that you
% find helpful for describing your paper; these are used to populate
% the "keywords" metadata in the PDF but will not be shown in the document
\icmlkeywords{Availability Poisoning, Unlearnable Examples, Data Poisoning, Shortcut Learning}

\vskip 0.3in
]

% this must go after the closing bracket ] following \twocolumn[ ...

% This command actually creates the footnote in the first column
% listing the affiliations and the copyright notice.
% The command takes one argument, which is text to display at the start of the footnote.
% The \icmlEqualContribution command is standard text for equal contribution.
% Remove it (just {}) if you do not need this facility.

\printAffiliationsAndNotice{}  % leave blank if no need to mention equal contribution
% \printAffiliationsAndNotice{\icmlEqualContribution} % otherwise use the standard text.

\begin{abstract}

Perturbative availability poisons (PAPs) add small changes to images to prevent their use for model training.
Current research adopts the belief that practical and effective approaches to countering PAPs do not exist. 
In this paper, we argue that it is time to abandon this belief.
We present extensive experiments showing that 12 state-of-the-art PAP methods are vulnerable to Image Shortcut Squeezing (ISS), which is based on simple compression.
For example, on average, ISS restores the CIFAR-10 model accuracy to $81.73\%$, surpassing the previous best preprocessing-based countermeasures by $37.97\%$ absolute.
ISS also (slightly) outperforms adversarial training and has higher generalizability to unseen perturbation norms and also higher efficiency.
Our investigation reveals that the property of PAP perturbations depends on the type of surrogate model used for poison generation, and it explains why a specific ISS compression yields the best performance for a specific type of PAP perturbation.
We further test stronger, adaptive poisoning, and show it falls short of being an ideal defense against ISS.
Overall, our results demonstrate the importance of considering various (simple) countermeasures to ensure the meaningfulness of analysis carried out during the development of PAP methods.
Our code is available at \url{https://github.com/liuzrcc/ImageShortcutSqueezing}.
\end{abstract}

\begin{figure}[ht!]
% \vskip 0.2in
\begin{center}
\centerline{\includegraphics[width=0.9\columnwidth]{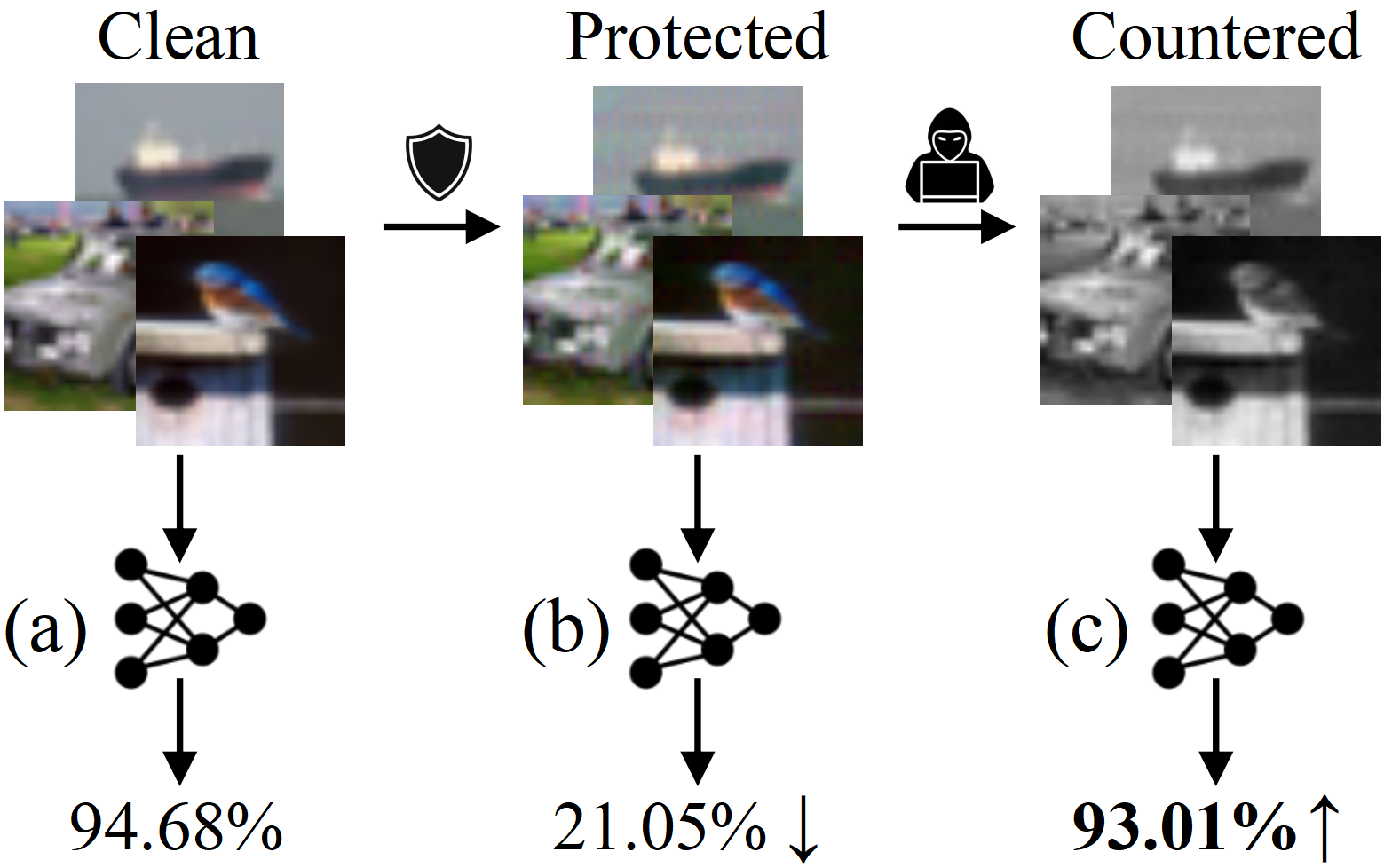}}
\caption{An illustration of our Image Shortcut Squeezing (ISS) for countering perturbative availability poisons (PAPs). The model accuracy is reduced by PAPs but is then restored by our ISS. Results are reported for EM~\cite{huang2021unlearnable} poisons on CIFAR-10.}
\label{fig:teaser}
\end{center}
\vskip -0.2in
\end{figure}

\section{Introduction}
\label{sec:intro}
The ever-growing amount of data that is easily available online has driven the tremendous advances of deep neural networks (DNNs)~\cite{schmidhuber2015deep, lecun2015deep, he2016deep, brown2020language}.
However, online data may be proprietary or contain private information, raising concerns about unauthorized use.
Perturbative availability poisons (PAPs) are recognized as a promising approach to data protection and recently a large number of PAP methods have been proposed that add perturbations to images which block training by acting as shortcuts~\cite{shen2019tensorclog, huang2021unlearnable, fowl2021adversarial,fowl2021preventing}.
As illustrated by Figure~\ref{fig:teaser} (a)$\rightarrow$(b), the high test accuracy of a DNN model is substantially reduced by PAPs.

Existing research has shown that PAPs can be compromised to a limited extent by preprocessing-based-countermeasures, such as data augmentations~\cite{huang2021unlearnable,fowl2021adversarial} and pre-filtering~\cite{fowl2021adversarial,anonymous2023selfensemble}.
However, a widely adopted belief is that no approaches exist that are capable of effectively countering PAPs.
Adversarial training (AT) has been proven to be a strong countermeasure~\cite{tao2021better,anonymous2023is}.
However, it is not considered to be a practical one, 
since it requires a large amount of computation and also gives rise to a non-negligible trade-off in test accuracy of the clean (non-poisoned) model~\cite{madry2018deep,zhang2019theoretically}. 
Further, AT trained with a specific $L_p$ norm is hard to generalize to other norms~\cite{tramer2019adversarial,laidlaw2020perceptual}.

In this paper, we challenge the belief that it is impossible to counter PAP methods both easily and effectively by demonstrating that they are vulnerable to simple compression.
First, we categorize 12 PAP methods into three categories with respect to the surrogate models they use during poison generation: 
slightly-trained~\cite{feng2019learning, huang2021unlearnable, yuan2021neural, fu2021robust, van2022generative}, fully-trained ~\cite{shen2019tensorclog, tao2021better, fowl2021adversarial, anonymous2023selfensemble}, and surrogate-free~\cite{wu2022one,yu2021indiscriminate, sandoval2022autoregressive}.
Then, we analyze perturbations/shortcuts that are learned with these methods and demonstrate that they are strongly dependent on features that are learned in different training stages of the model.
Specifically, we find that the methods using a slightly-trained surrogate model prefer \emph{low-frequency} shortcuts, while those using a fully-trained model prefer \emph{high-frequency} shortcuts.

Building on this new understanding, we propose Image Shortcut Squeezing (ISS), a simple, compression-based approach to countering PAPs.
As illustrated by Figure~\ref{fig:teaser} (b)$\rightarrow$(c), the low test accuracy of the poisoned DNN model is restored by our ISS to be close to the original accuracy.
In particular, grayscale compression is used to eliminate low-frequency shortcuts, and JPEG compression is used to eliminate high-frequency shortcuts.
We also show that our understanding of high vs. low frequency can also help eliminate surrogate-free PAPs~\cite{wu2022one,yu2021indiscriminate, sandoval2022autoregressive}.
Our ISS substantially outperforms previously studied data augmentation and pre-filtering countermeasures.
ISS also achieves comparable results to adversarial training and has three main advantages: 1) generalizability to multiple $L_p$ norms, 2) efficiency, and 3) low trade-off in clean model accuracy (see Section~\ref{sec: Evaluation in the Common Scenario} for details).

We further test the performance of ISS against potentially stronger PAP methods that are aware of ISS and can be adapted to it.
We show that they are not ideal against our ISS. 
Overall, we hope our study can inspire more meaningful analyses of PAP methods and encourage future research to evaluate various (simple) countermeasures when developing new PAP methods.

In sum, we make the following main contributions:

\begin{itemize}
    \item We identify the strong dependency of the perturbation frequency patterns on the nature of the surrogate model.
    Based on this new insight, we show that 12 existing perturbative availability poison (PAP) methods are indeed very vulnerable to simple image compression.
    
      \item We propose Image Shortcut Squeezing (ISS), a simple yet effective approach to countering PAPs.
      ISS applies image compression operations, such as JPEG and grayscale, to poisoned images for restoring the model accuracy.
      
      \item We demonstrate that ISS outperforms existing data augmentation and pre-filtering countermeasures by a large margin and is comparable to adversarial training but is more generalizable to multiple $L_p$ norms and more efficient.
    
    \item We explore stronger, adaptive poisons against our ISS and provide interesting insights into understanding PAPs, e.g., about the model learning preference of different perturbations.
\end{itemize}

\section{Related Work}
\label{related work}

\subsection{Perturbative Availability Poison (PAP)}
\label{sec: Availability Poisoning}

Perturbative availability poison (PAP) has been extensively studied.
TensorClog (TC)~\cite{shen2019tensorclog} optimizes the poisons by exploiting the parameters of a pre-trained surrogate to cause the gradient to vanish.
Deep Confuse (DC)~\cite{feng2019learning} collects the training trajectories of a surrogate classifier for learning a poison generator, which is computationally intensive.
Error-Minimizing (EM) poisons~\cite{huang2021unlearnable} minimizes the classification errors of images on a surrogate classifier with respect to their original labels in order to make them ``unlearnable examples''.
The surrogate is also alternatively updated to mimic the model training dynamics during poison generation.
Hypocritical (HYPO)~\cite{tao2021better} follows a similar idea to EM but uses a pre-trained surrogate rather than the above bi-level optimization.
Targeted Adversarial Poisoning (TAP)~\cite{fowl2021adversarial} also exploits a pre-trained model but minimizes classification errors of images with respect to incorrect target labels rather than original labels.

Robust Error-Minimizing (REM)~\cite{fu2021robust} improves the poisoning effects against adversarial training (with a relatively small norm) by replacing the normally-trained surrogate in EM with an adversarially-trained model.
Similar approaches~\cite{wang2021fooling, anonymous2023is} on poisoning against adversarial training are also proposed.
The usability of poisoning is also validated in scenarios requiring transferability~\cite{ren2022transferable} or involving unsupervised learning~\cite{he2022indiscriminate, zhang2022unlearnable}.

There are also studies focusing on revising the surrogate, e.g., Self-Ensemble Protection~\cite{anonymous2023selfensemble}, which aggregates multiple training model checkpoints, and NTGA~\cite{yuan2021neural}, which adopts the generalized neural tangent kernel to model the surrogate as Gaussian Processes~\cite{jacot2018neural}.
ShortcutGen (SG)~\cite{van2022generative} learns a poison generator based on a randomly initialized fixed surrogate and shows its efficiency compared to the earlier generative method, Deep Confuse.

Different from all the above methods, recent studies also explore surrogate-free PAPs~\cite{evtimov2021disrupting, yu2021indiscriminate, sandoval2022autoregressive}.
Intuitively, simple patterns, such as random noise~\cite{huang2021unlearnable} and semantics (e.g., MNIST-like digits)~\cite{evtimov2021disrupting}, can be used as learning shortcuts when they form different distributions for different classes.
Very recent studies also synthesize more complex, linear separable patterns to boost the poisoning performance based on sampling from a high dimensional Gaussian distribution~\cite{yu2021indiscriminate} and further refining it by introducing the autoregressive process~\cite{sandoval2022autoregressive}.
One Pixel Shortcut (OPS) specifically explores the model vulnerability to sparse poisons and shows that perturbing only one pixel is sufficient to generate strong poisons~\cite{wu2022one}.

In the domain of facial recognition, PAP methods, e.g., Fawkes~\cite{shan2020fawkes} and LowKey~\cite{cherepanova2021lowkey}, have also been studied.
However, their protection algorithms closely resemble the PAPs as discussed above. 
Specifically, Fawkes adopts a feature-layer loss similar to SEP and a robust surrogate model similar to REM, to boost transferability. 
LowKey adopts ensemble surrogate models similar to SEP and a pre-processing step similar to TAP, to boost transferability and imperceptibility.

In this paper, we evaluate our ISS against 12 representative PAP methods as presented above.
In particular, we consider poisons constrained by different $L_p$ norms.
Because of their technical similarity to two of the 12 approaches, we do not consider Fawkes and LowKey in our evaluation.

\subsection{PAP Countermeasures}
As mentioned in Section~\ref{sec:intro}, existing research has mainly relied on adversarial training (AT) for countering PAPs~\cite{tao2021better,anonymous2023is}.
However, AT is not practical due to the requirement of large computations and the non-negligible trade-off in test accuracy of the clean model~\cite{madry2018deep,zhang2019theoretically}. 
In addition, image preprocessing, e.g., data augmentations~\cite{huang2021unlearnable,fowl2021adversarial} and pre-filtering~\cite{fowl2021adversarial,anonymous2023selfensemble}, also show substantial effects but not comparable to AT.
In the domain of face recognition, countermeasures are also discussed but either require stronger assumptions or lack a concrete algorithm~\cite{radiya2021data}.
See more discussions in Appendix~\ref{sec: facial recognition}.

In this paper, we compare our ISS against existing countermeasures and particularly highlight its generalizability to unknown norms~\cite{tramer2019adversarial,laidlaw2020perceptual} and simplicity.

\subsection{Adversarial Perturbations and Countermeasures}
\label{sec:Adversarial Learning and Mitigation}

Simple image compressions, such as JPEG, bit depth reduction, and smoothing, are effective for countering adversarial perturbations based on the assumption that they are inherently high-frequency noise~\cite{dziugaite2016study, das2017keeping,xu2017feature}.
Other image transformations commonly used for data augmentations, e.g., resizing, rotating, and shifting, are also shown to be effective~\cite{xie2017mitigating,tian2018detecting,dong2019evading}.
However, such image pre-processing operations may be bypassed when the attacker is aware of them and then adapted to them~\cite{carlini2019evaluating}.
Differently, adversarial training (AT) is effective against adaptive attacks and is considered to be the most powerful defense so far~\cite{tramer2020adaptive}.
% AT has also been proven to be a principled defense against perturbative poisons~\cite{tao2021better}. 

Besides (training-time) data poisons, adversarial perturbations can also be used for data protection, but at inference time.
Related research has explored person-related recognition~\cite{oh2016faceless,oh2017adversarial,sattar2020body,rajabi2021practicality} and social media mining~\cite{larson2018pixel,li2019scene,liu2020exploring}.
An overview of inference-time data protection in images is provided by~\cite{orekondy2017towards}.

% Our ISS is based on compressions, and we particularly validate its compression effects in 
Our ISS is based on compression. 
We specifically evaluate its compression effects in Section~\ref{sec:relative strength of poisons}.

\section{Analysis of Perturbative Availability Poisons}
\label{sec:analysis}

\subsection{Problem Formulation}
\label{sec:Threat Model}

We formulate the problem of countering perturbative availability poisons (PAPs) in the context of image classification.
There are two parties involved, the data \emph{protector} and \emph{exploiter}.
The data protector poisons their own images to prevent them from being used by the exploiter for training a well-generalizable classifier.
Specifically, here the poisoning is achieved by adding imperceptible perturbations. 
The data exploiter is aware that their collected images may contain poisons and so apply countermeasures to ensure their trained classifier is still well-generalizable.
The success of the countermeasure is measured by the accuracy of the classifier on clean test images, and the higher, the more successful.

Formally stated, the protector aims to make a classifier $F$ generalize poorly on the clean image distribution $\mathcal{D}$, from which the clean training set $\mathcal{S}$ is sampled:
\vspace{-0.08cm}
% \begin{gather}
% \label{eq:outer}
%      \max_{\boldsymbol{\delta}} \,\, \mathbb{E}_{(\boldsymbol{x},y) \sim \mathcal{D}} \bigg[ \mathcal{L} \left( F(\boldsymbol{x}; \boldsymbol{\theta}(\boldsymbol{\delta})), y \right) \bigg] 
% \end{gather}
% \begin{gather}
% \label{eq:inner}
%      \text{s.t.} \,\, \boldsymbol{\theta}(\boldsymbol{\delta}) = \argmin_{\boldsymbol{\theta}}\sum_{(\boldsymbol{x}_i, y_i) \in \mathcal{S}} \mathcal{L}(F(\boldsymbol{x}_i + \boldsymbol{\delta}_i; \boldsymbol{\theta}(\boldsymbol{\delta})), y_i),
% \end{gather}
\begin{gather}
\label{eq:outer}
     \max_{\boldsymbol{\delta}} \,\, \mathbb{E}_{(\boldsymbol{x},y) \sim \mathcal{D}} \bigg[ \mathcal{L} \left( F(\boldsymbol{x}; \boldsymbol{\theta}'({\boldsymbol{\delta}})), y \right) \bigg] 
\end{gather}
\begin{gather}
\label{eq:inner}
     \text{s.t.} \,\, \boldsymbol{\theta}'({\boldsymbol{\delta}}) = \argmin_{\boldsymbol{\theta}({\boldsymbol{\delta}})}\sum_{(\boldsymbol{x}_i, y_i) \in \mathcal{S}} \mathcal{L}(F(\boldsymbol{x}_i + \boldsymbol{\delta}_i; \boldsymbol{\theta}({\boldsymbol{\delta}}), y_i),
\end{gather}
where $\boldsymbol{\theta}({\boldsymbol{\delta}})$ represents the parameters of the poisoned classifier, $F$, where $\boldsymbol{\delta}$ denotes the additive perturbations with $\epsilon$ as the $L_p$ bound. $\mathcal{L}(\cdot;\cdot)$ is the cross-entropy loss, which takes as input a pair of model output $F(\boldsymbol{x}_i; \boldsymbol{\theta})$ and the corresponding label $y_i$.

% In particular, $\boldsymbol{\theta}(\boldsymbol{\delta})$ represents the parameters of the poisoned classifier, where the perturbations $\boldsymbol{\delta}$ are added to the training images.

The exploiter aims to counter the poisons by applying a countermeasure $C$ to restore the model accuracy even when it is trained on poisoned data $\mathcal{P}$:
\begin{gather}
\label{eq:inner}
     \min_{\boldsymbol{\theta}}\sum_{(\boldsymbol{x}_i, y_i) \in \mathcal{P}} \mathcal{L}(F(C(\boldsymbol{x}_i + \boldsymbol{\delta}_i); \boldsymbol{\theta}), y_i).
\end{gather}

\subsection{Categorization of Existing PAP Methods}
We carried out an extensive survey of existing PAP methods, which allowed us to identify three categories of them regarding the type of their used surrogate classifiers.
These three categories are: Generating poisons 1) with a slightly-trained surrogate, 2) with a fully-trained surrogate, and 3) in a surrogate-free manner.
Table~\ref{tbl: Taxonomy of poisoning methods.} provides an overview of this categorization. 
In the first category, the surrogate is at its early training stage.
Existing methods in this category either fixes~\cite{yuan2021neural, van2022generative} or alternatively updates~\cite{feng2019learning, huang2021unlearnable,fu2021robust} the surrogate during optimizing the poisons.
In the second category, the surrogate has been fully trained.
Existing methods in this category fix the surrogate~\cite{shen2019tensorclog,tao2021better,fowl2021adversarial,anonymous2023selfensemble} but in principle, it may also be possible that the model is alternatively updated. 
In the third category, no surrogate is used but the poisons are synthesized by sampling from Gaussian distributions~\cite{yu2021indiscriminate,sandoval2022autoregressive} or optimized with a perceptual loss~\cite{wu2022one}.

\begin{figure*}[!t]
\begin{center}
\centerline{\includegraphics[width=\textwidth]{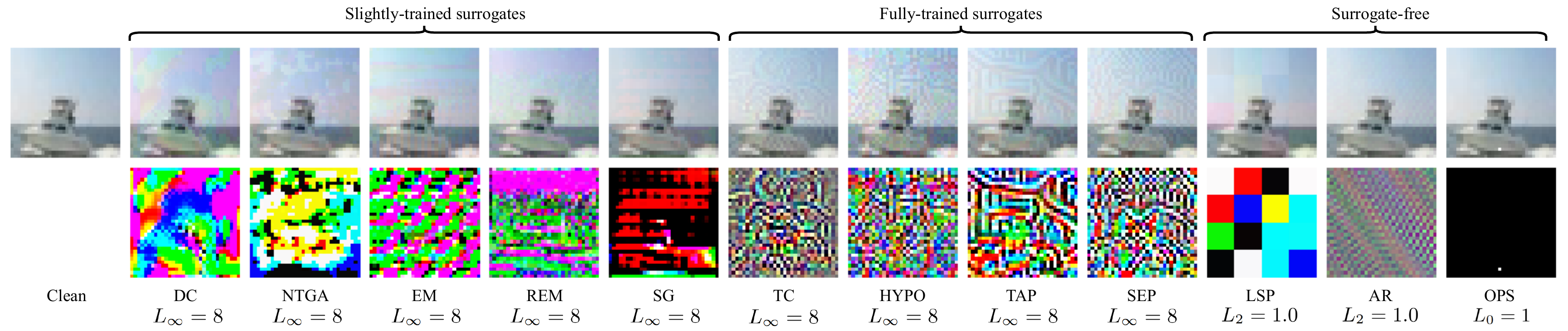}}
\caption{Poisoned CIFAR-10 images with corresponding perturbations.  Perturbations are re-scaled to $[0, 1]$ for visualization.}
\label{fig:examples}
\end{center}
\vspace{-0.5cm}

\end{figure*}

\begin{table}[!t]
\caption{Categorization of existing PAP methods.}
\label{tbl: Taxonomy of poisoning methods.}
\begin{center}
\begin{small}
% \begin{sc}
% \resizebox{\columnwidth}{!}{
\begin{tabular}{l|c}
\toprule
PAP Methods&Surrogate Model\\
\midrule
DC~\cite{feng2019learning}&\multirow{5}{*}{Slightly-Trained}\\
NTGA~\cite{yuan2021neural}&\\
EM~\cite{huang2021unlearnable}&\\
REM~\cite{fu2021robust}&\\
SG~\cite{van2022generative}&\\
 \hline
TC~\cite{shen2019tensorclog}&\multirow{4}{*}{Fully-Trained}\\
HYPO~\cite{tao2021better}&\\
TAP~\cite{fowl2021adversarial}&\\
SEP~\cite{anonymous2023selfensemble}&\\
 \hline
LSP~\cite{yu2021indiscriminate}&\multirow{3}{*}{Surrogate-Free}\\
AR~\cite{sandoval2022autoregressive}&\\
OPS~\cite{wu2022one}&\\
\bottomrule
\end{tabular}
% }
% \end{sc}
\end{small}
\end{center}
\vspace{-0.5cm}
\end{table}

\subsection{Frequency-based Interpretation of Perturbations}
\label{sec:fre}
Poisoned CIFAR-10 images and their corresponding perturbations for the 12 methods are visualized in Figure~\ref{fig:examples}.
As can be seen, the four methods that adopt a fully-trained surrogate tend to generate perturbations in patterns having a high spatial frequency.
This is consistent with the common finding in the adversarial example literature that adversarial perturbations are normally high-frequency~\cite{guo2018low}.
In contrast, the five methods that adopt a slight-trained surrogate exhibit spatially low-frequency patterns but large differences across color channels.

\begin{figure}[!t]
% \vskip 0.2in
\begin{center}
\centerline{\includegraphics[width=\columnwidth]{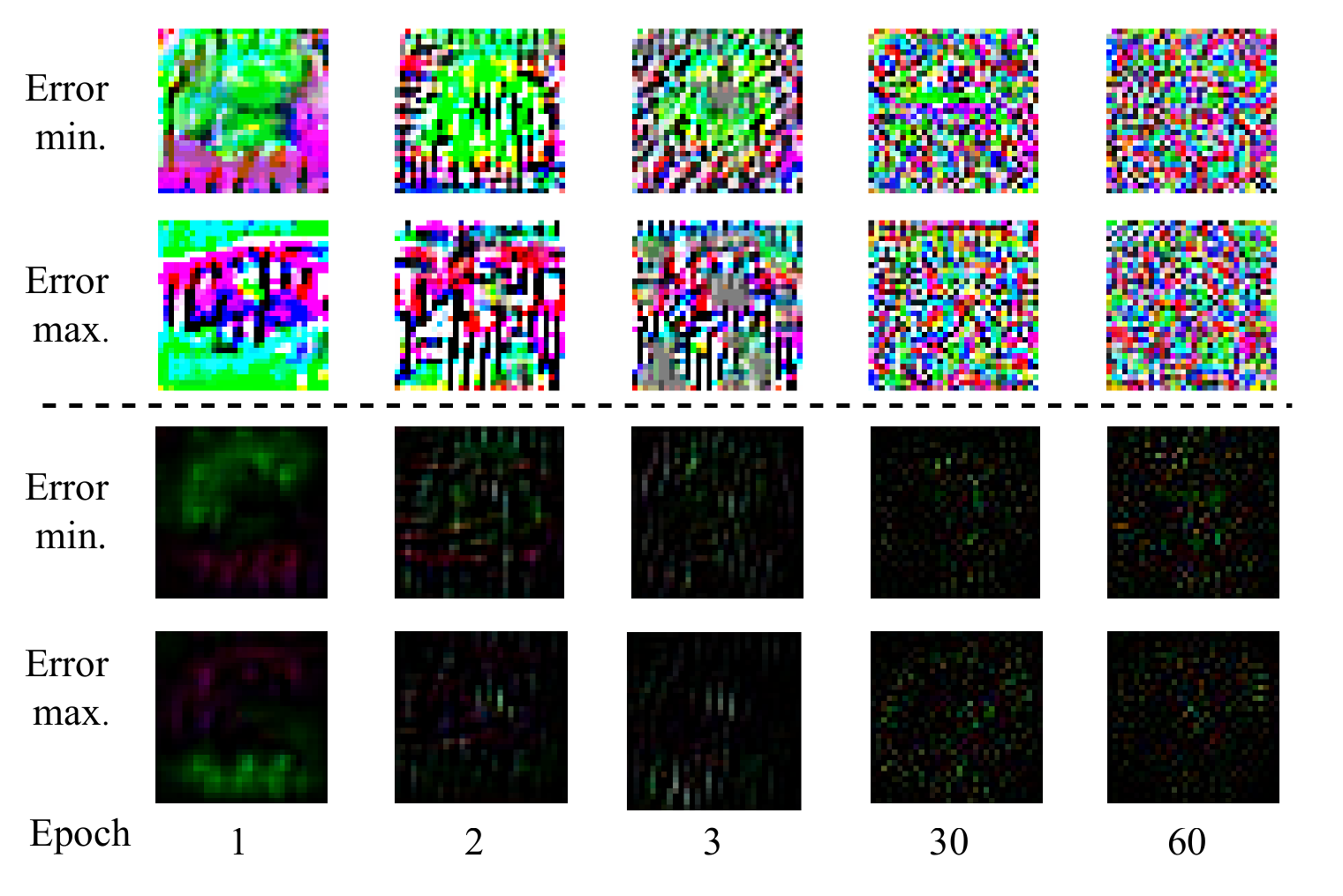}}
\caption{Perturbation visualizations for poisons generated using surrogate at its various training epochs. Perturbations with $L_{\infty} = 8$ (top) and $L_{2} = 1$ (bottom) are shown. Both the error minimizing and maximizing losses are considered. Perturbations at later epochs exhibit higher frequency.}
\label{fig:changeslinf}
\end{center}
% \vskip -0.2in
\vspace{-0.6cm}
\end{figure}

We hypothesize that the above phenomenon can be explained by the frequency principle~\cite{rahaman2019spectral,xu2019training,luo2019theory}, that
is, deep neural networks often fit target functions from low to high frequencies during training.
Accordingly, the poisons optimized against a slightly-trained model capture low-frequency patterns while those optimized against a fully-trained model capture high-frequency patterns~\cite{rahaman2019spectral,xu2019training,luo2019theory}.
In order to validate this hypothesis, we further try optimizing poisons using either the error-minimizing or error-maximizing loss against the surrogate at its various training epochs.
We visualize the resulting poisoned images and their corresponding perturbations in Figure~\ref{fig:changeslinf}.
As can be been, the spatial frequency of the perturbations gets increasingly higher as the surrogate goes to a later training epoch.

Different from those surrogate-based methods, the three surrogate-free methods have full control of the perturbation patterns they aim to synthesize. 
However, we notice that they still follow our frequency-based interpretation of perturbation patterns.
Specifically, the perturbations of LSP~\cite{yu2021indiscriminate} are uniformly upsampled from a Gaussian distribution and so exhibit patch-based low-frequency patterns.
On the other hand, the perturbations of AR~\cite{sandoval2022autoregressive} are generally based on sliding convolutions over the image and so exhibit texture-based high-frequency patterns.
OPS~\cite{wu2022one} perturbations only contain one pixel and so can be treated as an extreme case of high-frequency patterns.

\subsection{Our Image Shortcut Squeezing}
Based on the above new frequency-based interpretation, we propose Image Shortcut Squeezing (ISS), a simple, image compression-based countermeasure against PAPs.
We rely on different compression operations suitable for eliminating different types of perturbations.
Overall, a specific compression operation is applied to the $C(\cdot)$ in Eq.~\ref{eq:inner}.

For perturbations with low frequency but large differences across color channels, we use grayscale transformation to suppress such color differences.
We expect grayscale transformation to not sacrifice too much the test accuracy of a clean model because color information is known to contribute little to the DNNs' performance in differentiating objects~\cite{xie2018pre}.
For perturbations with high frequency, we follow existing research on eliminating adversarial perturbations to use common image compression operations, such as JPEG and bit depth reduction (BDR)~\cite{dziugaite2016study, das2017keeping,xu2017feature}.
We expect grayscale transformation to not sacrifice too much the test accuracy of a clean model because DNNs are known to be resilient to small amounts of  image compression, e.g., JPEG with a higher quality factor than 10~\cite{dodge2016understanding}.

\section{Experiments}
In this section, we evaluate our Image Shortcut Squeezing (ISS) and other countermeasures against 12 representative PAP methods.
We focus our experiments on the basic setting in which the surrogate (if it is used) and target models are the same and the whole training set is poisoned.
We also explore more challenging poisoning scenarios with unseen target models or partial poisoning (poisoning a randomly selected proportion or a specific class).

\subsection{Experimental Settings}
\label{sec:Experimental Settings}
\textbf{Datasets and models.} We consider three datasets: CIFAR-10~\cite{krizhevsky2009learning}, CIFAR-100~\cite{krizhevsky2009learning}, and a 100-class subset of ImageNet~\cite{deng2009imagenet}. 
If not mentioned specifically, on CIFAR-10 and CIFAR-100, we use 50000 images for training and 10000 images for testing.
For the ImageNet subset, we select 20\% images from the first 100 classes of the official ImageNet training set for training and all corresponding images in the official validation set for testing.
If not mentioned specifically, ResNet-18 (RN-18)~\cite{he2016deep} is used as the surrogate model and target model.
To study transferability, we consider target models with diverse architectures: ResNet-34~\cite{he2016deep}, VGG-19~\cite{simonyan2014very}, DenseNet-121~\cite{huang2017densely}, MobileNet-V2~\cite{sandler2018MobileNetV2}, and ViT~\cite{dosovitskiy2020image}.

\textbf{Training and poisoning settings.}
We train the CIFAR-10/100 models for 60 epochs and the ImageNet models for 100 epochs. 
We use SGD with a momentum of 0.9, a learning rate of 0.025, and cosine weight decay.
We adopt the torchvision transforms module for implementing Grayscale, JPEG, and bit depth reduction (BDR) in our Image Shortcut Squeezing (ISS).
We consider 12 representative existing poisoning methods as listed in Table~\ref{tbl: Taxonomy of poisoning methods.} under various $L_p$ norm bounds. 
A brief description of 12 methods can be found in Appendix~\ref{appendix: Brief Descriptions of Implemented Poisoning Methods}.
Specifically, we follow existing work and use $L_{\infty}=8$, $L_{2}=1.0$, and $L_{0}=1$.

\begin{table*}[ht!]
\caption{Clean test accuracy (\%) of models trained on CIFAR-10 poisons and with our Image Shortcut Squeezing (Gray and JPEG) vs. other countermeasures. Note that TC is known to not work well under small norms, e.g., our $L_{\infty}=8$~\cite{fowl2021adversarial}.
Hyperparameters for different countermeasures can be found in Appendix~\ref{appendix:Hyperparameters for Different Countermeasures}. 
}
\label{tbl:cifar10}
% \vskip 0.15in
\begin{center}
\begin{small}
% \begin{sc}
\resizebox{\textwidth}{!}{

\begin{tabular}{c|l|ccccccc|cccc|c}
\toprule
\multirow{2}*{Norm}&Poisons/Countermeasures&w/o&Cutout&CutMix&Mixup&Gaussian&Mean&Median&BDR&Gray&JPEG&Gray+JPEG&AT\\
\cmidrule{2-14}
% \midrule
&Clean (no poison) &94.68&95.10&95.50&95.01&94.17 &45.32 &85.94&88.65&92.41&85.38&83.79&84.99\\
\midrule
\multirow{9}*{$L_\infty=8$}&DC~\cite{feng2019learning}&16.30&15.14&17.99&19.39&17.21 &19.57 &15.82&61.10&\textbf{93.07}&81.84&83.09&78.00\\
&NTGA~\cite{yuan2021neural}  &42.46&42.07&27.16&43.03&42.84 &37.49 &42.91&62.50&\textbf{74.32}&69.49&69.86&70.05\\
&EM~\cite{huang2021unlearnable}&21.05&20.63&26.19&32.83&12.41&20.60 &21.70&36.46&\textbf{93.01}&81.50 &83.06&84.80\\ 
&REM~\cite{fu2021robust}  &25.44&26.54&29.02&34.48&27.44 &25.35 &31.57&40.77&\textbf{92.84 }&82.28&83.00&82.99\\
&SG~\cite{van2022generative}& 33.05&24.12&29.46&39.66&31.92& 46.87&49.53&70.14&\textbf{86.42}&79.49&79.21&76.38\\
\cline{2-14}
&TC~\cite{shen2019tensorclog}  &88.70&86.70&88.43&88.19& 82.58&72.25& 84.27& 84.85&79.75& 85.29&82.43& 84.55\\
&HYPO~\cite{tao2021better}  &71.54& 70.60& 67.54&72.54&72.46 &40.27 &65.53& 83.50&61.86&\textbf{85.45}&82.94&84.91\\
&TAP~\cite{fowl2021adversarial} &8.17&10.04&10.73&19.14&9.26&21.82&32.75&45.99&9.11&\textbf{83.87}&81.94&83.31\\
&SEP~\cite{anonymous2023selfensemble} &3.85&4.47& 9.41&15.59&3.96 &14.43 &35.65&47.43&3.57&\textbf{84.37}&82.18&84.12 \\

\midrule
\multirow{2}*{$L_2 = 1.0$}&LSP~\cite{yu2021indiscriminate} &19.07&19.87&20.89&26.99&19.25&28.85&29.85&66.19&82.47&\textbf{83.01}&79.05&84.59\\
&AR~\cite{sandoval2022autoregressive} &13.28&12.07&12.39&13.25&15.45&45.15&70.96&31.54&34.04&\textbf{85.15}&82.81&83.17\\
\midrule
$L_0 = 1$&OPS~\cite{wu2022one} &36.55&67.94&76.40&45.06&19.29&23.50&\textbf{85.16}&53.76&42.44&82.53&79.10&14.41\\
\bottomrule
\end{tabular}
}
% \end{sc}
\end{small}
\end{center}
% \vskip -0.1in
\end{table*}

\subsection{Evaluation in the Common Scenario}
\label{sec: Evaluation in the Common Scenario}
We first evaluate our ISS against 12 representative poisoning methods in the common scenario where the surrogate and target models are the same and the whole training dataset is poisoned.
Experimental results on CIFAR-10 shown in Table~\ref{tbl:cifar10} demonstrate that ISS can substantially restore the clean test accuracy of poisoned models in all cases.
Consistent with our new insight in Section~\ref{sec:fre}, grayscale yields the best performance in countering methods that rely on low-frequency perturbations with large color differences (see more results by other color compression methods on EM in Appendix~\ref{appendix: Color Channel Difference Mitigation Methods on EM}).
In contrast, JPEG and BDR are the best against methods that rely on high-frequency perturbations.
Additional results for other hyperparameters of JPEG and BDR in Table~\ref{tbl:jpeg bdr} of Appendix~\ref{appendix:Hyperparameters for Different Countermeasures} show that milder settings yield worse results.
In addition, we can also apply ISS without determining the specific poisons by directly using Gray+JPEG. 
The results demonstrate that this combination is globally effective against all 12 PAP methods, with only a small decrease in clean model accuracy.

Our ISS also outperforms other countermeasures. 
Specifically, data augmentations applied to clean models increase test accuracy but they are not effective against poisons. 
Image Smoothing is sometimes effective, e.g., median filtering performs the best against OPS as expected since it is effective against impulsive noise.
Adversarial training (AT) achieves comparable performance to our ISS for $L_{\infty}$ and $L_2$ norms but much worse performance for the $L_0$ norm. 
This verifies the higher generalizability of our ISS to unseen norms.
It is worth noting that the ISS training time is only $\frac{1}{7}$ of the AT training time on CIFAR-10.
The efficiency of our ISS becomes more critical when the dataset is larger and the image resolution is higher. 
Additional experimental results for $L_{\infty} = 16$ shown in Table~\ref{tbl:cifar10linf16} confirm the general effectiveness of our ISS.

We further conduct experiments on CIFAR-100 and ImageNet.
Note that for CIFAR-100, we only test the PAP methods that include CIFAR-100 experiments in their original work. 
For ImageNet, the poison optimization process is very time-consuming, especially for NTGA~\cite{yuan2021neural} and Deep Confuse~\cite{feng2019learning}.
Therefore, following the original work, these two methods are tested with only two classes.
Note that such time-consuming PAP methods are not good candidates for data protection in practice. 
Experimental results on CIFAR-100 and ImageNet shown in Table~\ref{tbl:cifar100} and Table~\ref{tbl:imagenet} confirm the general effectiveness of our ISS.

\begin{table}[!t]
\caption{Additional results on CIFAR-10 with larger perturbation norms: $L_{2}=2.0$ for LSP and $L_{\infty}=16$ for the rest.}
\label{tbl:cifar10linf16}
% \vskip 0.15in
\begin{center}
\begin{small}
% \begin{sc}
\resizebox{\columnwidth}{!}{

\begin{tabular}{l|cccccc|c}
\toprule
Poisons&w/o&Cutout&CutMix&Mixup&Gray&JPEG&AT\\
\midrule
Clean &94.68&95.10&95.50&95.01&92.41&88.65&84.99\\
\midrule
EM &16.33 &14.0 &13.41 &20.22 &60.85 &\textbf{63.44}&61.58\\ 
REM &24.89 &25.0 &22.85 &29.51 &42.85  &\textbf{76.59}& 80.14\\
HYPO &58.3 &54.22 &48.26 &57.27 &45.38  &\textbf{85.07}&84.90\\
TAP& 10.98&10.96& 9.46&17.97& 6.94&\textbf{84.19}&83.35\\
SEP&3.84 &8.90 &15.79 &9.27 &5.70 &\textbf{84.35} &84.07\\
\midrule
% LSP&11.97 &14.17 &17.98 &20.38 &\textbf{41.35} &40.02&80.22\\
LSP&19.07 &19.87 &20.89 &26.99 &\textbf{82.47} &83.01&84.59\\

\bottomrule
\end{tabular}
}
% \end{sc}
\end{small}
\end{center}
% \vskip -0.1in
\end{table}

% \vspace{-0.3cm}

\begin{table}[!t]
\caption{Additional results on CIFAR-100.}
\label{tbl:cifar100}
% \vskip 0.15in
\begin{center}
\begin{small}
% \begin{sc}
\resizebox{\columnwidth}{!}{

\begin{tabular}{l|ccccccc}
\toprule
Poisons&w/o&Cutout&CutMix&Mixup&Gray&JPEG\\
\midrule
Clean  &77.44&76.72&80.50&78.56&71.79&57.79\\
\midrule
EM    &7.25&6.70&7.03&10.68&\textbf{67.46}&56.01\\ 
REM&9.37&12.46&10.40&15.05&\textbf{57.27}&55.77\\
TC&57.52&60.56&59.19&59.77&47.93&58.94\\
TAP&9.00& 10.30&8.73&19.16&8.84&\textbf{83.77}\\
SEP&3.21&3.21&3.98&7.49&2.10&\textbf{58.18}\\
\midrule
LSP&3.06&4.43&6.12&5.61&44.62&\textbf{53.49}\\
AR&3.01&2.85&3.49&2.19&24.99&\textbf{57.87}\\
\midrule
OPS&23.78&\textbf{57.98}&56.03&22.71&32.62 &54.92\\
\bottomrule
\end{tabular}
}
% \end{sc}
\end{small}
\end{center}
% \vskip -0.1in
\end{table}

\begin{table}[!t]
\caption{Additional results on ImageNet subset. Following their original papers, NTGA and DC are tested with only two classes.}
\label{tbl:imagenet}
% \vskip 0.15in
\begin{center}
\begin{small}
% \begin{sc}
\resizebox{\columnwidth}{!}{

\begin{tabular}{l|cccccccc}
\toprule
Poisons&w/o&Cutout&CutMix&Mixup&Gray&JPEG\\
\midrule
Clean&62.04 &61.14&65.100&64.32&58.24&58.20\\
\midrule

EM&31.52&30.42&42.98&21.44&49.78&\textbf{49.88}\\ 
REM&11.12&11.62&12.50&17.62&\textbf{44.70}&18.16\\ 
TAP &24.64&23.00&18.72&28.62&24.30&\textbf{44.74} \\ 
LSP&26.32&27.64&17.22&2.5&\textbf{31.42}&30.78 \\ 
\midrule
NTGA&70.79&63.42&70.53&68.42&\textbf{83.42} &76.58\\ 
DC&65.00&-&-&-&\textbf{85.00}&74.00\\ 

\bottomrule
\end{tabular}
}
% \end{sc}
\end{small}
\end{center}
% \vskip -0.1in
\end{table}

\subsection{Evaluation in Challenging Scenarios}
\textbf{Partial poisoning.} In practical scenarios, it is common that only a proportion of the training data can be poisoned.
Therefore, we follow existing work~\cite{fowl2021adversarial, huang2020metapoison} to test such partial poisoning settings.
We poison a certain proportion of the training data and mix it with the rest clean data for training the target model.

Specifically, we test two partial poisoning settings: first, randomly selecting a certain proportion of the images, and, second, selecting a specific class.
In the first setting, as shown in Table~\ref{table partial cifar10}, the poisons are effective only when a very large proportion of the training data is poisoned.
For example, on average, even when 80\% of data are poisoned, the model accuracy is only reduced by about 10 \%.
In the second setting, we choose to poison all training samples from class \texttt{automobile} on CIFAR-10.
Table~\ref{tbl: table partial cifar10} demonstrates that almost all poisoning methods are very effective in the full poisoning setting.
In both settings, our ISS is effective against all PAP methods.

\begin{table}[ht!]
\caption{Clean test accuracy (\%) of CIFAR-10 target models under different poisoning proportions. TC is tested with $L_{\infty}=26$. 
}
\label{table partial cifar10}
% \vskip 0.15in
\begin{center}
\begin{small}
% \begin{sc}
\resizebox{\columnwidth}{!}{

\begin{tabular}{l|c|cccccc}
\toprule
Poisons&ISS &0.1&0.2&0.4&0.6&0.8&0.9\\
\midrule
\multirow{3}*{DC} & w/o&94.29 &94.26 &93.20 &91.66 &87.19&80.14  \\
&Gray &92.73 &92.57 &92.37 &91.51 &90.49&89.50  \\
&JPEG &84.89 &85.26 &84.43 &83.61 &83.02&82.69 \\
\midrule
\multirow{3}*{EM}&w/o &94.37 &93.63 &92.62 &91.07 &86.63&79.57  \\
&Gray &92.60 &92.62 &92.52 &92.23 &90.96&89.69  \\
&JPEG &84.61 &84.79 &84.96 &84.86 &84.93&84.40  \\
\midrule
\multirow{3}*{REM}&w/o&94.39&94.56&94.37&94.43&94.19&81.39 \\
&Gray&92.63&92.81&92.78&92.82&92.73&86.62 \\
&JPEG&84.64&85.53&84.82&85.37&85.38&82.44  \\
\midrule
\multirow{3}*{SG} &w/o &94.47 &94.40 &93.46 &91.21 &87.75&83.40 \\
&Gray&92.81 &92.65 &91.90 &90.65 &88.44&85.26  \\
&JPEG&84.94 &84.61 &84.11 &82.66 &80.76&79.38  \\
\midrule
\multirow{3}*{TC}&w/o &93.81 &94.09 &93.70 &93.59 &93.02&91.47  \\
&Gray &91.98 &92.38 &92.03 &91.96 &91.03&87.71 \\
&JPEG &85.24 &85.01 &85.23 &85.28 &85.23&84.37  \\
\midrule
\multirow{3}*{HYPO}&w/o &93.94 &94.43 &93.34 &92.56 &90.64&89.35 \\
&Gray &92.59 &92.39 &91.37 &90.06 &88.03&86.37 \\
&JPEG &85.61 &85.18 &85.39 &85.21 &85.25&85.10 \\
\midrule
\multirow{3}*{TAP}&w/o &94.09 &93.94 &92.75 &91.27 &88.42& 85.98\\
&Gray&92.62 &91.94 &90.73 &89.26 &85.93&83.18 \\
&JPEG &85.24 &84.42 &84.86 &84.98 &84.51&84.36\\
\midrule
\multirow{3}*{SEP}&w/o &94.12 &93.45 &92.76 &91.22 &87.82&85.01 \\
&Gray &92.57 &92.04 &91.09 &89.25 &86.31&82.95 \\
&JPEG &85.27 &85.27 &85.25 &84.71 &84.07&84.80  \\
\midrule
\multirow{3}*{LSP}&w/o&94.69 &94.42 &92.81 &91.38 &88.07&82.26 \\
&Gray &93.12 &92.56 &92.67 &92.20 &90.78&89.65  \\
&JPEG &85.01 &84.58 &84.88 &83.49 &83.27&81.67  \\
\midrule
\multirow{3}*{AR} &w/o&94.66 &94.38 &93.82 &91.80 &88.42&82.36\\
&Gray &92.85 &92.69 &92.53 &91.24 &89.88&85.35\\
&JPEG&85.37 &84.75 &85.35 &85.35 &85.07&87.27\\
\midrule
\multirow{3}*{OPS} &w/o&94.47&94.11&92.61&91.49&87.19&82.65 \\
&Gray&92.65&92.27&91.36&89.34&85.24&81.37 \\
&JPEG&84.75&84.88&84.55&83.98&82.87&81.33 \\
\bottomrule
\end{tabular}
}
% \end{sc}
\end{small}
\end{center}
% \vskip -0.1in
\end{table}

\begin{table}[ht!]
\caption{Partial poisoning for class \texttt{automobile} on CIFAR-10. TC is tested with $L_{\infty}=26$. 
}
\label{tbl: table partial cifar10}
% \vskip 0.15in
\begin{center}
\begin{small}
% \begin{sc}
% \resizebox{0.7\columnwidth}{!}{

\begin{tabular}{l|cccc}
\toprule
Poisons&w/o&Gray&JPEG&BDR\\
\midrule
DC&1.60	    &69.00	&88.30&52.20	\\
NTGA&51.70			&94.20&90.40&75.30\\
EM&0.10	   	&48.60	&94.30 &9.60\\
REM&0.80&34.40	&90.40	&2.50	\\
SG&27.75	&88.39	&78.59&70.05	\\
TC&0.50	    &0.20	&92.50&37.20	\\
HYPO&4.00	&3.00&94.90&56.80		\\
TAP&0.00		&0.10&93.90&38.10	\\
SEP&0.00	&0.00	&94.70	&15.50\\
LSP&67.30		&86.90&95.10&83.20	\\
AR&97.70&97.60	&94.60	&95.10	\\
OPS&28.90		&28.50	&93.60&72.10\\
\bottomrule
\end{tabular}
% }
% \end{sc}
\end{small}
\end{center}
% \vskip -0.1in
\end{table}

\begin{table}[!t]
\caption{Clean test accuracy (\%) of CIFAR-10 target models in the transfer setting. Note that AR, LSP, and OPS are surrogate-free. Four CNN models (ResNet-34, VGG-19, DenseNet-121, and MobileNet-V2) and one ViT are considered as the target model. TensorClog (TC) is tested with $L_{\infty}=26$.
}
\label{table transfer cifar10}
% \vskip 0.15in
\begin{center}
\begin{small}
% \begin{sc}
% \resizebox{\columnwidth}{!}{

\begin{tabular}{l|c|cccccc}
\toprule
Poisons & ISS &R34& V19&D121&M2& ViT\\
\midrule
\multirow{3}*{DC} &w/o&18.06 &16.59 &16.05 &17.81&24.09\\
&Gray&83.13 &80.32 &83.93 &78.78&44.83\\
&JPEG&82.64 &80.34 &83.38 &80.30&53.35\\

\midrule
\multirow{3}*{NTGA}&w/o&40.19 &47.13 &16.67 &40.75&31.82\\
&Gray&71.84 &76.89 &64.07 &62.28&58.25\\
&JPEG&67.00 &72.17 &73.76 &70.18&53.00\\
\midrule
\multirow{3}*{EM}&w/o&29.96 &34.70 &30.61 &30.10&18.84\\
&Gray&86.97 &87.03 &84.84 &82.81&63.28\\
&JPEG&84.21 &82.46 &84.86 &82.20&56.33\\
\midrule
\multirow{3}*{REM}&w/o&25.88 &29.04 &28.31 &24.08&32.22\\
&Gray&75.20 &77.99 &70.53 &66.21&63.00\\
&JPEG&82.35 &80.70 &81.74 &80.01&56.13\\
\midrule
\multirow{3}*{SG} &w/o&29.64 &48.5 &28.88 &30.75 &18.11\\
&Gray&86.53 &87.12 &86.07 &81.34 &42.22\\
&JPEG&79.57 &77.78 &79.77 &75.87 &56.27\\
\midrule
\multirow{3}*{TC}&w/o&87.71 &85.47 &78.04 &78.51&69.86\\
&Gray&78.48 &75.14 &66.72 &62.39&61.86\\
&JPEG&84.56 &82.66 &83.95 &82.60&55.51\\
\midrule
\multirow{3}*{HYPO} &w/o &80.64&81.59&81.48&78.27&67.49\\
&Gray&75.25&76.65&74.29&69.81&53.02\\
&JPEG&85.55&83.39&85.03&83.95&55.17\\
\midrule

\multirow{3}*{TAP}&w/o&7.89& 8.59& 8.64 &10.02&41.32\\
&Gray&9.38 &11.51& 8.77& 8.29& 42.49\\
&JPEG&84.42 &81.95 &84.28 &82.24&57.35\\
\midrule
\multirow{3}*{SEP}&w/o&3.11& 6.70& 4.41& 5.29 &25.56\\
&Gray&4.00& 5.40& 4.20& 4.70 &22.23\\
&JPEG&84.64 &83.38 &84.55 &83.25 &56.94\\
\midrule

\multirow{3}*{LSP}&w/o&15.98 &17.39 &19.79 &17.32&26.65\\
&Gray&71.10 &82.11 &73.06 &70.61&53.36\\
&JPEG&79.57 &78.72 &79.66 &76.79&60.41\\
\midrule
\multirow{3}*{AR} &w/o&21.31 &19.78 &13.54 &16.08&22.91\\
&Gray&70.54 &76.92 &67.35 &62.01& 53.22\\
&JPEG&85.62 &83.95 &85.46 &83.50&54.88\\

\midrule
\multirow{3}*{OPS} &w/o&37.06 &36.3 &40.03 &27.35&30.25\\
&Gray&44.29 &42.21 &38.32 &38.71&21.77\\
&JPEG&82.84 &80.70 &82.83 &80.42&62.93\\

\bottomrule
\end{tabular}
% }
% \end{sc}
\end{small}
\end{center}
% \vskip -0.1in
\end{table}

\textbf{Transferability to unseen models.} In realistic scenarios, the protector may not know the details of the target model.
In this case, the transferability of the poisons is desirable. 
Table~\ref{table transfer cifar10} demonstrates that all PAP methods achieve high transferability to diverse model architectures and our ISS is effective against all of them. 
It is also worth noting that there is no clear correlation between the transferability and the similarity between the surrogate and target models.
For example, transferring from ResNet-18 to ViT is not always harder than to other CNN models. 

\subsection{Adaptive Poisons to ISS}
\label{sec: adaptive poisoning}
In the adversarial example literature, image compression operations can be bypassed when the attacker is adapted to them~\cite{shin2017jpeg, carlini2019evaluating}.
Similarly, we evaluate strong adaptive poisons against our ISS using two PAP methods, EM ($L_{\infty}$) and LSP ($L_{2}$).
We assume that the protector can be adapted to grayscale and/or JPEG in our ISS.
Specifically, for EM, we add a differentiable JPEG compression module~\cite{shin2017jpeg} and/or a differentiable grayscale module into its bi-level poison optimization process.
For LSP, we increase the patch size to 16$\times$16 to decrease high-frequency features so that JPEG will be less effective, and we make sure the pixel values are the same for three channels to bypass grayscale.

Table~\ref{tbl:adaptive} demonstrates that for EM, the adaptive grayscale poisons are effective against grayscale, but adaptive JPEG-10 noises are not effective against JPEG.
As hinted by~\cite{shin2017jpeg}, using an ensemble of JPEG with different quality factors might be necessary for better adaptive poisoning.
We also implement BPDA~\cite{athalye2018obfuscated} with the same JPEG quality factor (i.e., JPEG-10) and find that our ISS still ensures a very high model accuracy, i.e., 83.70 \%.
For LSP, we observe that even though adaptive LSP is more effective against the combination of JPEG and grayscale than the other two individual compressions, it is insufficient to serve as a good adaptive protector.
On the other hand, adaptive LSP also fails against the model without ISS, indicating that the additional operations (grayscale and larger patches) largely constrain its poisoning effects.

Given that the protector may have full knowledge of our ISS, we believe that better-designed adaptive poisons can bypass our ISS in the future.

\begin{table}[!t]
\caption{Clean test accuracy (\%) of four different target models under EM poisoning and its adaptive variants on CIFAR-10. Results are reported for $L_{\infty}=8$ and Table~\ref{tbl:adaptive_16} in Appendix reports results of EM for $L_{\infty}=16$, which follow the same pattern.}
\label{tbl:adaptive}
% \vskip 0.15in
\begin{center}
\begin{small}
% \begin{sc}
% \resizebox{0.9\columnwidth}{!}{

\begin{tabular}{l|cccc|c}
\toprule
Poisons&w/o	&Gray &JPEG&G\&J&Ave.\\
\midrule
EM       &21.05&93.01&81.50&83.06&69.66\\
EM-Gray    &17.81&\textbf{16.60}&76.71&74.16&\textbf{46.32} \\
EM-JPEG &\textbf{17.11}&89.18&83.11&82.85&68.06\\
EM-G\&J &48.93&46.29&\textbf{69.48}&\textbf{66.26}&57.74\\
\midrule
LSP   &19.07 &82.47 &83.01&79.05&65.90 \\
LSP-G\&J &93.01&90.34&84.38&\textbf{82.13}&87.47\\
\bottomrule
\end{tabular}
% }
% \end{sc}
\end{small}
\end{center}
% \vskip -0.1in
\end{table}

\subsection{Further Analyses}
\label{sec:relative strength of poisons}

\begin{figure}[!t]
% \vskip 0.2in
\begin{center}
\centerline{\includegraphics[width=0.8\columnwidth]{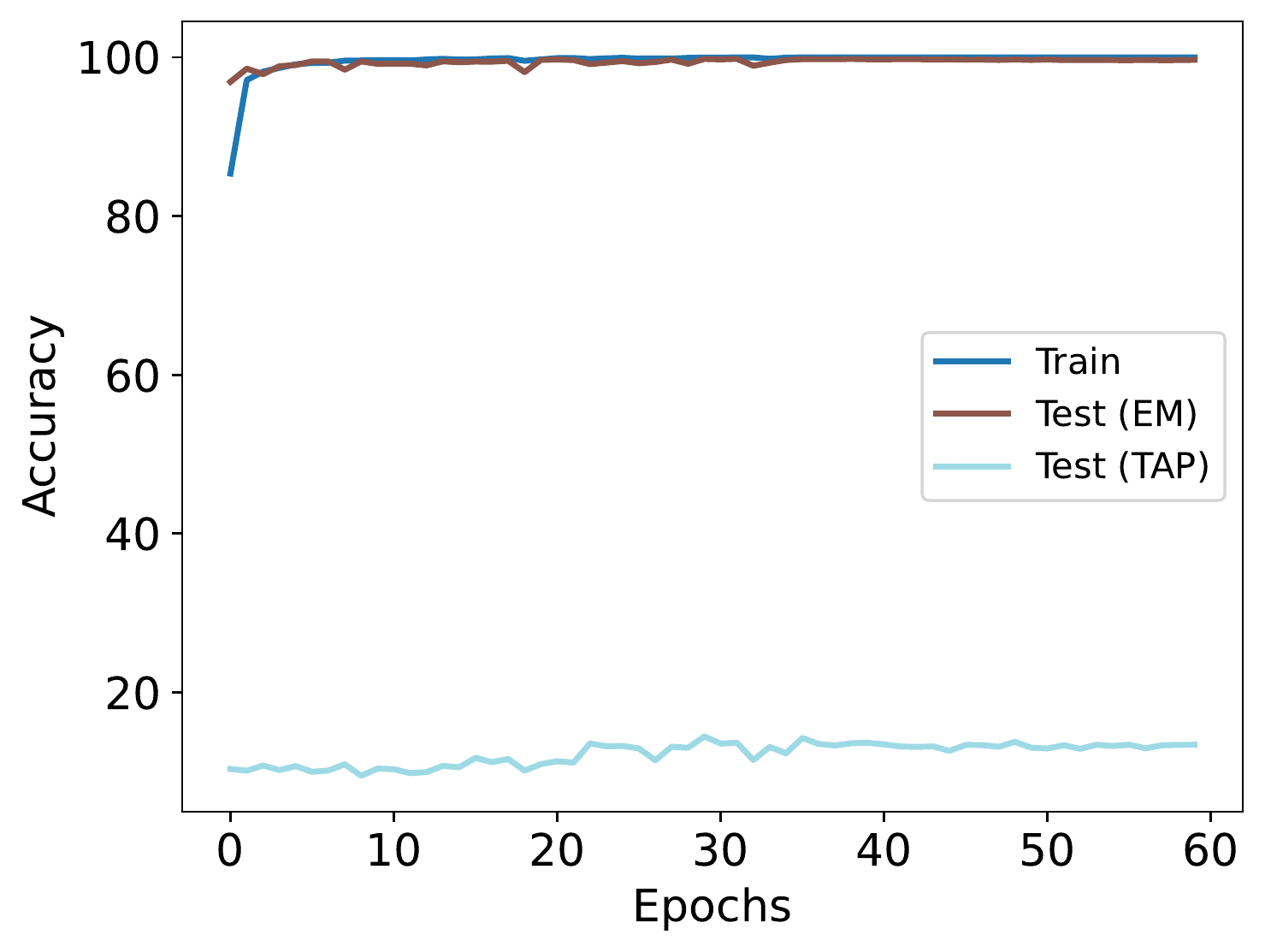}}
\caption{Relative model preference of different poisons.
}
\label{fig:stronger}
\end{center}
\vskip -0.3in
\end{figure}

\textbf{Working Mechanism of ISS.}
Here we illustrate the working mechanism of our ISS by ensuring that the poisons are the exact factor that is used by the poisoned model for prediction.
To this end, we follow~\cite{fowl2021adversarial} to use poisoned images to both train and test the model. In this case, if the test accuracy (on poisoned images) is high, it demonstrates that the perturbations in the poisoned images are actually learned by the model. 
In addition, we also train and test the model on poisoned images but differently, the testing (poisoned) images are pre-processed using our ISS. In this case, if the test accuracy (on poisoned images) decreases, it demonstrates that ISS can suppress the perturbations at inference time.
The results in Table~\ref{tbl:why} validate our hypotheses.

\begin{table*}[!t]
\caption{Test accuracy (\%) on clean, poisoned, and ISS-preprocessed poisoned test sets of models that are trained on different poisons.}
\label{tbl:why}
% \vskip 0.15in
\begin{center}
\begin{small}
% \begin{sc}
% \resizebox{\columnwidth}{!}{

\begin{tabular}{l|cccccccccccc}
\toprule
Test/ Poisons	&DC	&NTGA	&EM	&REM	&SG	&TC	&HYPO	&TAP	&SEP	&LSP	&AR	&OPS \\
\midrule
Clean	&17.96	&-	&16.77	&26.04	&37.50	&87.86	&73.06	&11.63	&4.91	&15.29	&16.37	&17.50\\
Poisoned	&97.20	&97.85	&99.85	&99.97	&96.72	&93.79	&99.98	&100.0	&99.99	&100.0	&99.94	&99.83\\
Poisoned+ISS	&11.17	&24.86	&11.89	&20.68	&25.39	&24.16	&16.68	&11.33	&10.13	&10.06	&13.83	&12.61\\

\bottomrule[1pt]
\end{tabular}
% }
% \end{sc}
\end{small}
\end{center}
\vskip -0.1in
\end{table*}

\textbf{Relative Model Preference of different poisons.} We explore the relative model preference of low-frequency vs. high-frequency poisons.
This scenario is practically interesting because the same online data might be poisoned by different methods.
Inspired by the experiments on the model preference of MNIST vs. CIFAR data in~\cite{shah2020pitfalls}, we simply add up the EM and TAP perturbations for each image.
The perturbation norm is doubled accordingly.
For example, for perturbations with $L_{\infty} = 8$, the composite perturbations range from $-16$ to $16$.
We train a model (using the original image labels) on the composite perturbations of EM and TAP and test it on either EM or TAP perturbations.

As shown in Figure~\ref{fig:stronger}, the model converges fast and reaches a high test accuracy on EM but not on the TAP.
It indicates that TAP perturbations are less preferred than EM perturbations by the model during training.

\textbf{ISS for a combination of different types of poisons.}
We create poisons by combining the two well-known low-frequency and high-frequency methods, i.e., EM and TAP. Specifically, we take the average of the perturbations of these two methods. As shown in Table~\ref{tbl:combine}, our ISS is still effective against this combination.

\begin{table}[!t]
\caption{Clean test accuracy (\%) of models trained on CIFAR-10 poisons that is a combination of low-frequency poison EM and high-frequency TAP. }
\label{tbl:combine}
% \vskip 0.15in
\begin{center}
\begin{small}
% \begin{sc}
% \resizebox{0.7\columnwidth}{!}{

\begin{tabular}{l|ccc}
\toprule

Poisons/ISS&w/o	&Gray &JPEG\\
\midrule
EM	&21.05	&93.01	&81.50\\
TAP	&8.17	&9.11	&83.87\\
EM+TAP	&36.07	&18.93	&84.62\\

\bottomrule
\end{tabular}
% }
% \end{sc}
\end{small}
\end{center}
% \vskip -0.1in
\end{table}

\textbf{ISS for both training and testing.}
Our ISS only applies to the training data for removing the poisons.
However, in this case, it may cause a possible distribution shift between the training and test data. 
Here we explore such a shift by comparing ISS with another variant that applies compression to both the training and test data.
Table~\ref{tbl:traintest-test} demonstrates that in most cases, these two versions of ISS do not lead to substantial differences.

\begin{table}[!t]
\caption{Clean test accuracy (\%) for ISS (Gray and JPEG), which applies compression only to training data or to both training and test data (denoted with suffix-TT).}
\label{tbl:traintest-test}
% \vskip 0.15in
\begin{center}
\begin{small}
% \begin{sc}
% \resizebox{0.9\columnwidth}{!}{

\begin{tabular}{l|cc|cc}
\toprule

Poisons&Gray-TT	&Gray	&JPEG-TT	&JPEG	\\
\midrule
Clean	&\textbf{92.62}	&92.41	&79.56	&\textbf{85.38}	\\
\midrule
DC	&83.79	&\textbf{93.07}	&79.41	&\textbf{81.84}	\\
NTGA	&65.42	&\textbf{74.32}	&62.84	&\textbf{69.49}	\\
EM	&90.75	&\textbf{93.01}	&78.96	&\textbf{81.50}	\\
REM	&73.38	&\textbf{92.84}	&79.39	&\textbf{82.28}	\\
% SG	&50.08	&\textbf{50.38}	&69.5	&\textbf{75.34}	\\
SG	& \textbf{88.26}	&86.42	& 72.96	&\textbf{ 79.49}	\\
TC	&\textbf{76.41}	&75.88	&79.42	&\textbf{83.69}	\\
HYPO	&\textbf{75.20}	&61.86	&79.63	&\textbf{85.60}	\\
TAP&	\textbf{9.53}&	9.11	&78.65	&\textbf{83.87}	\\
SEP	&2.93 &\textbf{3.57}&79.28 &\textbf{84.37}\\
LSP	&\textbf{76.23}	&75.77	&68.73	&\textbf{78.69}	\\
AR	&68.95	&\textbf{69.37}	&79.26	&\textbf{85.38}	\\
OPS	&\textbf{46.53}	&42.44	&76.87	&\textbf{82.53}\\

\bottomrule
\end{tabular}
% }
% \end{sc}
\end{small}
\end{center}
\vskip -0.2in
\end{table}

\section{Conclusion and Outlook}
In this paper, we challenge the common belief that there are no practical and effective countermeasures to perturbative availability poisons (PAPs).
Specifically, we show that 12 state-of-the-art PAP methods can be substantially countered by Image Shortcut Squeezing (ISS), which is based on simple compression.
ISS outperforms other previously studied countermeasures, such as data augmentations and adversarial training.
Our in-depth investigation leads to a new insight that the property of PAP perturbations depends on the type of surrogate model used during poison generation. 
We also show the ineffectiveness of adaptive poisons to ISS.
We hope that further studies could consider various (simple) countermeasures during the development of new PAP methods.

For future work, on the countermeasure side, we would further improve ISS on the trade-off between its effectiveness and the decrease of clean model accuracy by exploring other (simple) accuracy-preserving operations.
In addition, to achieve a countermeasure that is more effective against unknown poisons, it would be promising to explore more advanced combination strategies of operations or conduct automatic attack identification and then apply attack-specific operations.
On the protection side, we encourage future work to develop effective (adaptive) protection methods against our ISS and other potential countermeasures.

\section*{Acknowledgements}
We are grateful to Alex Kolmus, Dirren van Vlijmen, Rui Wen, Tijn Berns, Tom Heskes, Twan van Laarhoven, and the anonymous reviewers for discussion and feedback on this work.
We thank Shutong Wu, Jie Ren, and Hao He for providing source codes.
Part of this work was carried out on the Dutch national e-infrastructure with the support of SURF Cooperative.

\newpage
\newpage\clearpage
\bibliography{ref}
\bibliographystyle{icml2023}

%%%%%%%%%%%%%%%%%%%%%%%%%%%%%%%%%%%%%%%%%%%%%%%%%%%%%%%%%%%%%%%%%%%%%%%%%%%%%%%
%%%%%%%%%%%%%%%%%%%%%%%%%%%%%%%%%%%%%%%%%%%%%%%%%%%%%%%%%%%%%%%%%%%%%%%%%%%%%%%
% APPENDIX
%%%%%%%%%%%%%%%%%%%%%%%%%%%%%%%%%%%%%%%%%%%%%%%%%%%%%%%%%%%%%%%%%%%%%%%%%%%%%%%
%%%%%%%%%%%%%%%%%%%%%%%%%%%%%%%%%%%%%%%%%%%%%%%%%%%%%%%%%%%%%%%%%%%%%%%%%%%%%%%
\newpage

\clearpage
\appendix
\onecolumn

\section{Brief Descriptions of Implemented PAP Methods}
\label{appendix: Brief Descriptions of Implemented Poisoning Methods}
\begin{itemize}
    \item \textbf{Deep Confuse (DC)}~\cite{feng2019learning}: Perturbations are generated from a U-Net~\cite{ronneberger2015u} on CIFAR-10 and encoder-decoder model on two-class ImageNet. The generators are trained on the output of a pseudo-updated classifier, where the classification model is first trained on clean data and then trained on adversarial data to update the generator. We use the implementation from the official GitHub repository.\footnote{\url{https://github.com/kingfengji/DeepConfuse}}
        
    \item \textbf{Neural Tangent Generalization Attacks (NTGA)}~\cite{yuan2021neural} (target model-agnostic): NTGA uses Neural Tangent Kernels to approximate target networks and then leverages the approximation to generate perturbations. We use the poisons provided in the official GitHub repository.\footnote{\url{https://github.com/lionelmessi6410/ntga}}
    
    \item \textbf{Error-Minimizing perturbations (EM)}~\cite{huang2020metapoison}: Bi-level optimizing error-minimizing perturbations after certain steps of training on perturbed samples that are from the last iteration. The surrogate model is trained on-the-fly with perturbed training samples. We use the implementation from the official GitHub repository.\footnote{\url{https://github.com/HanxunH/Unlearnable-Examples/}}
    
    \item \textbf{Robust Error-Minimizing perturbations (REM)}~\cite{fu2021robust}: Same as EM, but the model is adversarially trained and the perturbations generation is equipped with expectation over transformation technique (EOT)~\cite{athalye2018synthesizing}. We use the implementation from the official GitHub repository.\footnote{\url{https://github.com/fshp971/robust-unlearnable-examples/}}
    
    \item \textbf{Shortcut generator (SG)}~\cite{van2022generative}: Perturbations are generated from a ResNet-like encoder-decoder model from~\cite{naseer2019cross}. Different from another generative poisoning Deep Confuse, the discriminator model is randomly initialized without training. We use the CIFAR-10 poisons (version `SG') provided by the authors by private communication.
    
    \item \textbf{TensorClog (TC)}~\cite{shen2019tensorclog}: A second-order derivative with respect to training data is calculated to iteratively optimize the perturbations to minimize the gradients of model loss with respect to the weights of model layers. We use the implementation from the official GitHub repository for poisons ($L_{\infty} = 26$) on CIFAR-10\footnote{\url{https://github.com/JC-S/TensorClog_Public}}. We also use the implementation from \url{https://github.com/lhfowl/adversarial_poisons} for poisons ($L_{\infty} = 8, 16$) on CIFAR-10.

    \item \textbf{Hypocritical perturbations (HYPO)}~\cite{tao2021better}: Similar to EM, but the error-minimizing perturbations are generated on a pre-trained surrogate model which is trained on clean data. We use the implementation from the official GitHub repository.\footnote{\url{https://github.com/TLMichael/Delusive-Adversary}}
        
    \item \textbf{Targeted Adversarial Poisoning (TAP)}~\cite{fowl2021adversarial}: Targeted adversarial examples by PGD~\cite{madry2018deep} and Spatial Transformer Networks (STN) module~\cite{jaderberg2015spatial}. The poisoning target labels are different from the original labels, but the target labels are the same for poisoning images whose clean versions are from the same class. We use the implementation from the official GitHub repository.\footnote{\url{https://github.com/lhfowl/adversarial_poisons}}

    \item \textbf{Self-Ensemble Protection (SEP)}~\cite{anonymous2023selfensemble} SEP ensembles intermediate checkpoints when training on the clean training set to create perturbations. SEP is currently the state-of-the-art protection on CIFAR-10.  We use the implementation from the official GitHub repository.\footnote{\url{https://github.com/Sizhe-Chen/SEP}}
        
    \item \textbf{Linear separable Synthetic Perturbations (LSP)}~\cite{yu2021indiscriminate}: Linearly separable Gaussian samples are listed by order and then up-scaled to the size of the image. Perturbations that are sampled from the same Gaussian are added to the same class. We use the implementation from the official GitHub repository.\footnote{\url{https://github.com/dayu11/Availability-Attacks-Create-Shortcuts/}}
    
    \item \textbf{AutoRegressive poisoning (AR)}~\cite{sandoval2022autoregressive} Autoregressive process generates perturbations that CNN favors during training.  We use the CIFAR-10 poisons provided by the authors in the official GitHub repository.\footnote{\url{https://github.com/psandovalsegura/autoregressive-poisoning}}
    
    \item \textbf{One Pixel Shortcut (OPS)}~\cite{wu2022one}: OPS generates one pixel shortcut by searching the pixel that creates the most significant mean pixel value change for all images from one class. The perturbations are \emph{dataset-dependent}. We use the implementation from the official GitHub repository.\footnote{\url{https://github.com/cychomatica/One-Pixel-Shotcut}}

\end{itemize}

\section{Hyperparameters for Different Countermeasures}
\label{appendix:Hyperparameters for Different Countermeasures}

If not explicitly mentioned, we use JPEG with a quality factor of 10 and bit depth reduction (BDR) with 2 bits.
For grayscale compression, we use the torchvision implementation\footnote{\url{https://pytorch.org/vision/stable/generated/torchvision.transforms.Grayscale}} where the weighted sum of three channels are first calculated and then copied to all three channels. 
For adversarial training (AT), PGD-10 is used with a step size of $\frac{2}{255}$, where the model is trained on CIFAR-10 for 100 epochs.
We use a kernel size of $3\time3$ for both median, mean, and Gaussian smoothing (with a standard deviation of $0.1$).

% \section{Additional Results on Adaptive Poisons}
% \label{appendix:Additional Results on Adaptive Poisons}

\begin{table*}[!t]
\caption{JPEG with different quality factors and BDR with different bit depth.}
\label{tbl:jpeg bdr}
% \vskip 0.15in
\begin{center}
\begin{small}
% \begin{sc}
\resizebox{\textwidth}{!}{

\begin{tabular}{lcccccccccccc}
\toprule
\multirow{2}*{Poisons}&\multirow{2}*{w/o}&\multicolumn{5}{c}{JPEG Compression}&\multicolumn{5}{c}{Bit depth reduction}\\
&&10&30&50&70&90&2&3&4&5&6\\
\midrule
Clean (no poison)&94.68&85.38&89.49&90.80&91.85&93.06&88.65&92.22&93.45&94.46&94.55\\
\midrule
DC~\cite{feng2019learning} &16.30&81.84&79.35&69.69&58.53&34.79&61.10&27.03&17.34&16.42&15.11&\\
NTGA~\cite{yuan2021neural}  &42.46&69.49&66.83&64.28&60.19&53.24&62.58&53.48&47.30&44.39&43.29\\
EM~\cite{{huang2021unlearnable}}&21.05&81.50&70.48&54.22&42.23&21.98&36.46&24.99&22.57&21.54&20.60 \\
REM~\cite{fu2021robust} &25.44&82.28&77.73&71.19&63.39&37.89&40.77&28.81&28.39&25.38&26.49\\
SG~\cite{van2022generative} &33.05&79.49&77.15&74.49&73.03&70.76&69.32&58.03&47.33&31.67&31.56\\
HYPO~\cite{tao2021better} &71.54&85.45&89.14&90.16&88.10&70.66&83.17&80.33&76.91&73.22&72.05\\ 
TAP~\cite{fowl2021adversarial} &8.17&83.87&84.82&77.98&57.45&11.97&45.99&18.29&14.16&8.590&7.38\\
SEP~\cite{anonymous2023selfensemble}&3.85&84.37&87.57&82.25&59.09&8.06&43.48&10.01&7.89&4.99&3.66\\
LSP~\cite{yu2021indiscriminate} &15.09&78.69&42.11&33.99&29.19&26.66&48.27&29.56&25.14&16.88&14.27\\
AR~\cite{sandoval2022autoregressive}&13.28&85.15&89.17&86.11&80.01&54.41&31.54&12.64&11.66&9.96&12.99\\
OPS~\cite{wu2022one} &36.55&82.53&79.01&68.58&59.81&53.02&53.76&48.46&46.79&38.44&42.27\\
\bottomrule
\end{tabular}
}
% \end{sc}
\end{small}
\end{center}
% \vskip -0.1in
\end{table*}

\begin{table}[!t]
\caption{Clean test accuracy (\%) of target models under EM poisoning and its adaptive variants on CIFAR-10. Results are reported for $L_{\infty}=16$.}
\label{tbl:adaptive_16}
% \vskip 0.15in
\begin{center}
\begin{small}
% \begin{sc}
\resizebox{0.5\columnwidth}{!}{

\begin{tabular}{l|cccc|c}
\toprule

Poisons&w/o	&Gray &JPEG&G\&J&Ave.\\
\midrule
EM      &19.32&80.60&84.32&82.12&66.59\\
EM-Gray   &10.01&12.14&50.14&52.07& 31.09\\
EM-JPEG &21.63&64.83&68.21&81.22&58.97\\
EM-G\&J &19.71&22.68&28.94&30.51&25.46\\
\bottomrule
\end{tabular}
}
% \end{sc}
\end{small}
\end{center}
% \vskip -0.1in
\end{table}

\section{Color Channel Difference Mitigation Methods on EM}
\label{appendix: Color Channel Difference Mitigation Methods on EM}

We show that grayscale compression is a special case where the weighted sum of different channels are used.
Table~\ref{tbl:cmean} demonstrates that other approaches that reduce color channel differences can also be applied to counter poisons.

\section{PAP Countermeasures in Facial Recognition}
\label{sec: facial recognition}
In the domain of facial recognition,~\cite{radiya2021data} propose two countermeasures against two PAP methods, Fawkes~\cite{shan2020fawkes} and LowKey~\cite{cherepanova2021lowkey}. Their first countermeasure is based on robust training via data augmentation and assumes that an additional clean pre-trained model is available to the data exploiter.
In contrast, our work explores robust training, via adversarial training, but does not assume the exploiter has access to additional (clean) data and model. 
Table~\ref{tbl:RT} demonstrates that models trained by their robust training on one type of poison would not generalize to others, limiting the effectiveness of the robust data augmentation against PAPs.

Their second countermeasure is more conceptual, which is to ``wait for better facial recognition systems to be developed in the future.'' This method clearly depends on the potential progress of future models and obviously cannot act as an effective solution at this moment.
In contrast, our ISS requires no change to the existing model but only applies pre-processing operations.

\begin{table}[!t]
\caption{Clean test accuracy (\%) of CIFAR-10 when train and test on different poisons.}
\label{tbl:RT}
% \vskip 0.15in
\begin{center}
\begin{small}
% \begin{sc}
\resizebox{\columnwidth}{!}{

\begin{tabular}{l|cccccccccccc}
\toprule[1pt]
Train / Test	&DC	&NTGA	&EM	&REM	&SG	&TC	&HYPO	&TAP	&SEP	&LSP	&AR	&OPS \\
\midrule
DC	&\textbf{97.20}	&18.76	&17.24	&19.39	&19.01	&18.05	&18.39	&17.86	&18.02	&14.09	&17.43	&17.79\\
NTGA	&20.74	&\textbf{97.85}	&28.06	&33.24	&37.98	&33.10	&38.71	&30.54	&32.69	&36.07	&40.35	&38.41\\
EM	&14.49	&15.65	&\textbf{99.85}	&20.31	&19.57	&15.79	&15.08	&14.52	&14.06	&12.58	&16.99	&16.22\\
REM	&21.34	&22.44	&23.31	&\textbf{99.97}	&24.13	&24.42	&24.94	&22.48	&23.69	&25.89	&25.93	&26.41\\
SG	&31.99	&36.63	&31.53	&26.89	&\textbf{96.71}	&33.41	&35.87	&29.55	&30.49	&39.59	&39.43	&36.43\\
TC	&64.27	&67.41	&50.04	&61.93	&75.41	&\textbf{93.79}	&61.42	&61.57	&72.03	&76.81	&73.81	&87.28\\
HYPO	&63.58	&67.51	&71.55	&67.50	&71.63	&21.59	&\textbf{99.98}	&1.59	&0.56	&70.60	&73.69	&72.31\\
TAP	&9.15	&10.68	&10.81	&10.61	&11.28	&18.96	&9.10	&\textbf{100.00}	&9.16	&10.61	&12.48	&11.65\\
SEP	&4.50	&5.28	&4.65	&5.11	&4.86	&5.53	&6.45	&4.58	&\textbf{99.99}	&5.06	&4.99	&5.50\\
LSP	&16.35	&18.33	&25.47	&22.07	&17.79	&16.95	&21.08	&20.03	&19.36	&\textbf{100.00}	&16.35	&15.31\\
AR	&14.06	&16.15	&10.42	&15.60	&20.53	&17.64	&16.15	&14.28	&14.55	&16.82	&\textbf{99.94}	&12.85\\
OPS	&13.11	&18.26	&16.78	&13.53	&16.69	&12.29	&14.04	&15.98	&12.39	&16.79	&17.45	&\textbf{99.83}\\
\bottomrule[1pt]
\end{tabular}
}
% \end{sc}
\end{small}
\end{center}
% \vskip -0.1in
\end{table}

\begin{table}[!t]
\caption{Clean test accuracy (\%) of ResNet-18 trained on EMs that are pre-processed by another three channel-wise color suppression methods. C-mean calculates the mean value and copies the mean to three channels. R-3/G-3/B-3 copies the values from the red/green/blue channel to three channels. 
}
\label{tbl:cmean}
% \vskip 0.15in
\begin{center}
\begin{small}
% \begin{sc}
\resizebox{0.5\columnwidth}{!}{

\begin{tabular}{l|ccccc}
\toprule[1pt]
Methods&C-Mean&R-3&G-3&B-3&Gray\\
Acc.&91.83&86.60&86.73&87.91&93.01\\
\bottomrule[1pt]
\end{tabular}
}
% \end{sc}
\end{small}
\end{center}
% \vskip -0.1in
\end{table}
% You can have as much text here as you want. The main body must be at most $8$ pages long.
% For the final version, one more page can be added.
% If you want, you can use an appendix like this one, even using the one-column format.
%%%%%%%%%%%%%%%%%%%%%%%%%%%%%%%%%%%%%%%%%%%%%%%%%%%%%%%%%%%%%%%%%%%%%%%%%%%%%%%
%%%%%%%%%%%%%%%%%%%%%%%%%%%%%%%%%%%%%%%%%%%%%%%%%%%%%%%%%%%%%%%%%%%%%%%%%%%%%%%

\end{document}